\documentclass[preprint,showpacs,superscriptaddress,preprintnumbers,amsmath,
amssymb]{revtex4}
\usepackage{graphicx}
\usepackage{subfigure}
\usepackage{dcolumn}
\usepackage{bm}
\begin{document}


\title{A study of the $p\,d \, \rightarrow \, p \,d\,\eta$ reaction}
\author{N.J.Upadhyay}
\email{neelam@physbu.mu.ac.in}
\affiliation{Department of Physics, University of Mumbai, Vidyanagari, 
Mumbai - 400 098, India}
\author{K.P.Khemchandani}
\email{kanchan@ific.uv.es}
\affiliation{Department of Physics, University of Mumbai, Vidyanagari, 
Mumbai - 400 098, India}
\affiliation{Departamento de F\'isica Te\'orica and IFIC, 
Centro Mixto Universidad de
Valencia-CSIC Institutos de Investigaci\'on de Paterna,
Aptd. 22085, 46071 Valencia, Spain}
\author{B.K.Jain}
\email{brajeshk@gmail.com}
\affiliation{Department of Physics, University of Mumbai, Vidyanagari, 
Mumbai - 400 098, India}
\author{N.G.Kelkar}
\email{nkelkar@uniandes.edu.co}
\affiliation{Departamento de Fisica, Universidad de los Andes, Cra.1E No.
18A-10, Santafe de Bogota, Colombia}
\date{\today}
\begin{abstract}
A study of the $p\,d \, \rightarrow \, p \,d\,\eta$ reaction 
in the energy range where the recent data from Uppsala are available, 
is done in the two-step model of $\eta$ production including the final 
state interaction. The $\eta -d$ final state interaction is incorporated 
through the solution of the Lippmann Schwinger equation using an   
elastic scattering matrix element, $T_{\eta\,d\,\rightarrow\,\eta\,d}$, 
which is required to be half off-shell. It is written in a factorized 
form, with an  off-shell form factor multiplying an on-shell part given 
by an effective range expansion up to the fourth power in momentum. The 
parameters of this expansion have been taken from an existing 
recent relativistic Faddeev equation solution for the $\eta NN$ 
system corresponding to different $\eta-N$ scattering amplitudes.
Calculations have also been done using few body equations 
within a finite rank approximation (FRA) to generate 
$T_{\eta\,d\,\rightarrow\,\eta\,d}$. The $p-d$ final state interaction
is included in the spirit of the Watson-Migdal prescription by 
multiplying the matrix element by the inverse of the Jost function. 
The $\eta-d$ interaction is found to be dominant in the region 
of small invariant $\eta -d$ mass, $M_{\eta d}$.
The $p-d$ interaction enhances the cross section in the whole region 
of $M_{\eta d}$, but is larger for large $M_{\eta d}$. We find nearly 
isotropic angular distributions of the proton and the deuteron in the
final state. All the above observations are in agreement with data. 
The production mechanism for the entire range of the existing data on
the $p\,d\,\rightarrow\,p\,d\,\eta$ reaction seems to be dominated 
by the two-step model of $\eta$ production.
\end{abstract}


\maketitle

\section{Introduction}
The current great interest in the $\eta $-nucleus interaction exists 
because of the attractive nature of the $\eta-N$ interaction 
in the s-wave \cite{bhale}, and the consequent possibility of the 
existence of quasi-bound, virtual or resonant $\eta $-nucleus states 
\cite{hailiu}. The exact nature of these states, of course, depends 
upon the precise knowledge of the $\eta-N$ scattering matrix at low 
energies. As the $\eta$ is a highly unstable meson (lifetime 
$\sim 10^{-18} s$), this precise information is 
difficult to obtain directly. It can only be obtained from the eta 
producing reactions through the final state interaction. With this 
motivation, starting with the early experiments near threshold at Saclay 
on the $p\,d\,\rightarrow \,^3He \,\eta$ and the $p\,d\,\rightarrow \,
p\,d\,\eta$ reactions, measurements have been carried out near threshold 
and beyond at J\"ulich and Uppsala using the COSY and Celsius rings 
respectively. In this series of experiments, the recent 
data on the $p\,d \, \rightarrow \, p \,d\,\eta$ reaction 
using the Wasa/Promice setup at the Celsius storage
ring of the Svedberg laboratory, Uppsala are 
thematically complete and cover the excess energy, Q,
($Q\, = \,\sqrt{s}\,-\,m_\eta\,-\,m_p\,-\,m_d$) ranging
from around threshold to 107 MeV. The data \cite{bilger} 
(integrated over other variables) include the 
invariant mass distribution over the whole excess energy range 
for the $\eta -d$, $\eta -p$ and $p-d$ systems and angular 
distributions for the proton, deuteron and the eta meson. 
Like the $p\,d\,\rightarrow \,^3He \,\eta$ reaction, 
the (inclusive) $\eta -d$ invariant mass distribution exhibits a large 
enhancement near threshold and hence appears promising to 
study the $\eta -d$ interaction. The $\eta -p$ and $p-d$ invariant mass 
distributions do not show any such enhancement. 
All observed angular distributions are nearly isotropic. 

Like in our earlier studies on the $p\,d\,\rightarrow \,^3He \,\eta$ 
reaction our primary aim in this paper is to investigate the above 
mentioned data on the $\eta -d$ invariant mass distribution to obtain a 
better understanding of the $\eta -N$ as well as the $\eta -d$ interaction. 
We speculate, from our experience on the study of the $p\,d\,\rightarrow 
\,^3He \,\eta$ reaction \cite{we3}, that in the region of low $\eta -d$ 
relative energy this set of data will be mainly determined by the 
$\eta -d$ interaction, though the three-body nature of the final state 
may introduce some uncertainty in this conclusion. 
   
We present a study of the $p\,d\,\rightarrow \,p\,d\,\eta$ reaction which 
includes the effect of the final state interaction. 
We have investigated two possible diagrams for the production mechanism:
the direct mechanism and the two step process of $\eta$ production. 
The direct mechanism proceeds via an intermediate $p\,n\,\rightarrow\,d\,\eta$
reaction with one of the nucleons in the deuteron as a spectator. 
The $\eta$ meson in the two step model is produced in two steps, namely, 
 $p\,p\,\rightarrow \,d \,\pi^+$ and $\pi^+\,N\,\rightarrow\,\eta\,N$, hence
involving the participation and sharing of the transferred momentum by three 
nucleons. The two step model for $\eta$ production was first used in 
\cite{lage} and the data on the $p\,d\,\rightarrow\,^3He\,\eta$ reaction was 
well explained. The vertices at the two steps have been described by the 
corresponding off-shell $T$-matrices. The $T$-matrix for 
$\pi^+\,N\,\rightarrow\,\eta\,N$ is taken from a coupled channel calculation 
\cite{bhale}, and that for $p\,p\,\rightarrow \,d \,\pi^+$ is obtained 
from the SAID program provided by the authors of Ref. \cite{arndt}.

The final state interaction between the $\eta$ and the deuteron is 
explicitly incorporated through an $\eta-d$ $T$-matrix, $T_{\eta d}$.  
This $T$-matrix, which is required to be half-off-shell, 
is described in two ways. One choice involves taking a ``factorized form" 
which is given by an off-shell form factor 
multiplied by an on-shell part given by an effective range expansion up 
to the fourth power in momentum. The parameters of this expansion have been 
taken from an existing recent relativistic Faddeev 
equation solution for the $\eta NN$ system \cite{gar} corresponding to
different $\eta-N$ scattering amplitudes. 
The off-shell form factor will be described in the next sections and is 
chosen to have a form without any adjustable parameters. 
The second prescription involves solving few body equations within the finite
rank approximation (FRA) to obtain $T_{\eta d}$. 
This approach has been used in literature for the $\eta -d$, 
$ -^3$He and $ -^4$He systems \cite{rakit}. 
We perform calculations for both the prescriptions using different models of 
the elementary coupled channel $\eta$-nucleon $T$-matrix which characterize 
them.

The interaction between the $\eta $ meson and the proton in the 
final state, to a certain extent is contained implicitly in our 
calculations. This is due to the fact that we describe the 
$\pi^+\,N\,\rightarrow\,\eta\,N$ vertex by a $T$-matrix, which has been 
modeled to include the $\eta-N$ interaction. 
This off-shell $T$-matrix treats the $\pi\,N$, $\eta\,N$
and $\pi\,\Delta$ channels in a coupled channel formalism
\cite{bhale} and  reproduces the experimental data on this 
reaction very well.

The effect of $p-d$ final state interaction (FSI) is incorporated in 
the spirit of the Watson-Migdal FSI prescription \cite{watmig}, in which  
our model $p\,d \, \rightarrow \, p \,d\,\eta$ 
production amplitude is multiplied by a factor which incorporates the 
FSI between the proton and the deuteron. This factor is taken to be 
the frequently used \cite{jost,gold,shyam,dks} inverse 
Jost function, $[J(p)]^{-1}$, where $p$ is the relative 
$p-d$ momentum. The assumption implicit in this 
approximation that the mechanism for the primary reaction be short 
ranged is very well fulfilled in the $\eta $-production reactions. 
The momentum transfer in these reactions near threshold is around 
700 MeV/c. We include FSI for both doublet ($^2S_{1/2}$) and quadruplet 
($^4S_{3/2}$) $p-d$ states.    

The $\eta$-nucleon $T$-matrix, which characterizes our calculations, 
is not precisely known. Recent theoretical works on the 
$n\,p\,\rightarrow\,d\,\eta$ reaction 
\cite{penya} conclude that
the data on this reaction can be reproduced with the strength of the real
part of the $\eta$-nucleon
scattering length ranging between 0.42 and 0.72 fm. 
In our earlier work on the $p\,d\,\rightarrow\,^3He\,\eta$ reaction 
\cite{we3}, we found a good agreement with data, with the real part of the 
scattering length taken to be around 0.75 fm. This value was also found 
to be in agreement 
with the $n\,p\,\rightarrow\,d\,\eta$ data in a $K$-matrix calculation of the 
final state $\eta-d$ interaction in \cite{wycech}.
The same authors as in \cite{wycech}, recently performed a fit 
to a wide variety of data which includes the 
$\pi\,N\,\to\,\pi\,N$, $\pi\,N\,\to\,\eta\,N$, $\gamma\,N\,\to\,\pi\,N$ and 
$\gamma\,N\,\to\,\eta\,N$ reactions and gave their best fit 
value of the $\eta$-nucleon scattering length, $a_{\eta N}$ to be 
$(0.91, 0.27)$ fm \cite{green}. The $\eta-d$ effective range 
parameters are given in \cite{gar} for $a_{\eta N}$ up to $(1.07\,,\,0.26)$ fm.
Hence, in the present work we perform calculations with different models 
of the $\eta-N$ interaction, which correspond to three different values of 
the $\eta-N$ scattering length, ranging from $a_{\eta N}\,=\,(0.42\,,\,0.34)$ 
fm to $(1.07\,,\,0.26)$ fm.

We find that the cross sections calculated using the two-step model 
and the above inputs for the final state interaction 
reproduce most of the features of the experimental data reasonably well. 

A theoretical effort to understand the Uppsala data 
\cite{bilger} was made earlier by Tengblad {\it et. al.} \cite{wilkin}.
In \cite{wilkin} the contribution of three different diagrams, namely, the 
pick-up (a direct one-step mechanism of $\eta$ production), the impulse 
approximation and the two-step mechanism (here the $\eta$ meson is produced in 
two steps via the $p\,p\,\rightarrow\,\pi^+\,d$ and 
$\pi^+\,N\,\rightarrow\,\eta\,N$ reactions) to the cross section for the 
$p\,d\,\rightarrow\,p\,d\,\eta$ reaction is determined. The authors in 
\cite{wilkin} conclude that the impulse approximation is in general 
negligible as compared to the other two diagrams, 
the two-step mechanism is dominant in the near threshold region and the
contribution of the pick-up diagram (referred to as the direct mechanism in 
the present work) increases with energy and matches the two-step contribution 
at an excess energy of $Q = 95$ MeV. The latter conclusions regarding the
contributions of the two step and pick up diagrams are in contrast to the 
findings of the present work as well as to existing literature on 
similar kind of reactions. We note here that the authors in \cite{wilkin} 
do not include the final state interaction in their calculations in any way. 
They treat the kinematics and the dependence of the pion propagator 
(appearing in the two step model) on the Fermi 
momenta in an approximate way. The $T$-matrices which enter as an 
input to the two step model are simply extracted from experimental 
cross sections and are hence not proper off-shell $T$-matrices.
As a result of the above approximations, the authors in \cite{wilkin} do not 
reproduce the observed enhancement in the $\eta -d$ invariant mass 
distribution near threshold, and unlike the observed isotropic distributions, 
find anisotropy in their calculated angular distributions. 
 
The contribution from the direct mechanism (or the so-called pick-up diagram of
\cite{wilkin}) to the total cross sections is found to be 
about four orders of magnitude smaller than the two-step contribution
at threshold in the present work. 
The one-step contribution does increase with energy
(as also found in \cite{wilkin}), however, even at the highest energy
for which data is available ($T_p = 1096$ MeV) it remains two orders
of magnitude smaller than that due to the two-step model. This is in
contrast to the observations in \cite{wilkin}, where the two processes give
comparable contributions at high energies.
The difference of orders of magnitude between the two processes can be  
understood as a result of the large 
momentum transfer, $q$, in the one-step process. This $q$, which
is very large in the threshold region ($\sim 840$ MeV/c) continues 
to be large even at high energies. For example, it is $\sim 600$ MeV/c 
even at the highest beam energy of 1096 MeV.
This finding of ours is very similar to the previous studies of the
reactions involving high momentum transfer. 
For example, as mentioned above too, in \cite{lage}, for the 
$pd\rightarrow ^3He \eta $ reaction 
up to 2.5 GeV beam energy, the authors  
comment that the one-step cross 
sections underestimate the data by more than two orders of magnitude. In yet 
another calculation \cite{kond} of the cross section for the 
$pd\rightarrow ^3$He $X$ reaction (where $X = \eta , \eta^{\prime}, \omega, 
\phi$) the two-step model was found to describe
the data on these reactions up to $3$ GeV quite well.
In \cite{kama}, in connection with the $p\,d\,\rightarrow\,^3{\rm H}_\Lambda\,
 K^+$ reaction, 
the authors claim that for a beam energy of 
1 - 3 GeV, the one-step mechanism predicts 2 to 3 orders of magnitude smaller 
cross sections as compared to the two-step mechanism. The cross sections 
obtained from the one-step model, in Ref. \cite{wilkin} are, however, reported 
to be only one order of magnitude less than those due to the two-step model 
at threshold and comparable to the two-step ones at high energies. 

In the next section, we describe the details of the formalism. 
In the subsequent sections we present and discuss the results and finally 
the conclusions.

\section{The Formalism}
The differential cross section for the $p\, d\, \rightarrow \,p \,d \,\eta$ 
reaction, in the center of mass, can be written as,
\begin{equation}\label{sig}
d\sigma\,=\,{m_p^2\,m_d^2 \over 2\,(2\,\pi)^5\,s\,|\vec{k_p}|}\,\,
d\Omega_{p^\prime}\,\,|\vec{k_{p^\prime}}|\,\,dM_{\eta\,d}\,\,
|\vec{k}_{\eta\,d}|\,\,d\Omega_{\eta\,d}\,\,{1 \over 6}\,\,
\langle\,|T|^2\,\rangle
\end{equation}
where $\sqrt s$ is the total energy in the center of mass and $\vec{k_p}$  
and $\vec{k_{p^\prime}}$ are the proton momenta in the 
initial and final states respectively. $M_{\eta\,d}$ denotes the invariant 
mass of the $\eta -d$ system and $\vec{k}_{\eta\,d}$ and $\Omega_{\eta\,d}$ 
denote, in the $\eta -d$ center of mass, the $\eta $ momentum and its solid 
angle, respectively. $\Omega_{p^\prime}$ represents the solid angle of the 
outgoing proton. Angular brackets around $|T|^2$ in Eq. (\ref{sig}) represent 
the sum over the final and initial spins.

The $T$-matrix, which includes the interaction between the $\eta$ and 
the deuteron is given by
\begin{equation}\label{one}
T = \langle \, \psi_{\eta d}(\vec{k_{\eta d}}), \,\vec{k_{p^\prime}}; 
\,m_{p^\prime}, \,m_{d^\prime}\,|\,T_{p d \rightarrow p d \eta} \,|\, 
\vec{k_p},\, \vec{k_d}\,(=\,-\,\vec{k_p});\,m_p, \,m_d  \, \rangle
\end{equation}
where the spin projections for the proton and 
the deuteron in the
initial and final states have been labeled as $m_p, \,m_d,\,m_{p^\prime},$ 
and $m_{d^\prime}$ respectively. $T_{p\,d\,\rightarrow\,p\,d\,\eta}$ is the 
production operator. 

The wave function of the interacting $\eta -d$ in the final state has been 
represented as $\psi_{\eta d}(\vec{k_{\eta d}})$. 
In terms of the elastic $\eta -d$ scattering 
$T$-matrix, $T_{\eta d}$, it is written as 
\begin{equation}\label{psi}
\langle\,\psi^-_{\eta d}\,| = \langle \,\vec{k_{\eta d}}\,| + 
\int {d\vec{q} \over (2\pi)^3} {\langle \,\vec{k_{\eta d}}\,|\, 
T_{\eta d}\,|\,\vec{q}\, \rangle \over E(k_{\eta d})\,-\,E(q) + 
i\epsilon} \langle \,\vec{q}\,|
\end{equation}
The second term here represents the scattered wave. It has 
two parts originating from the 
principal-value and the  delta-function part of the propagator in the 
intermediate state. Physically they represent the off-shell and the on-shell 
scattering between the $\eta $ and the deuteron. The on-shell part can be 
shown to be roughly proportional to the $\eta -d$ momentum and hence 
dominant at higher energies. The relative contribution of these terms in 
our case would be determined after we      
substitute the above expression for $\psi_{\eta d}(\vec{k_{\eta d}})$ 
in Eq. (\ref{one}). 
We then get 
\begin{equation}\label{tfull}
T = \langle \, \vec{k_{\eta d}}, \,\vec{k_{p^{\prime}}}; 
\,m_{p^\prime}, \,m_{d^\prime}
\,|\,T_{p d \rightarrow p d \eta} \,|\, \vec{k_p},\, \vec{k_d} (= - \vec{k_p});
\,m_p, \,m_d  \, \rangle
\end{equation}
$$+\sum\limits_{m_{2^\prime}} \int {d\vec{q} \over (2\pi)^3} {\langle \,
\vec{k_{\eta d}}; m_{d^\prime} \,|\, T_{\eta d}\,|\,\vec{q}; m_{2^\prime}\,
\rangle \over E(k_{\eta d}) - E(q) + i\epsilon} \langle \,\vec{q}\,, 
\vec{k_{p^\prime}}; m_{2^\prime}, m_{p^\prime}\,|
\,T_{p d \rightarrow p d \eta} \,|\, \vec{k_p},\, \vec{k_d};
\,m_p, m_d \rangle$$
It can be seen that the $T_{\eta d}$ here appears as a half-off-shell 
$T$-matrix.
\begin{figure}
\includegraphics[width =9cm]{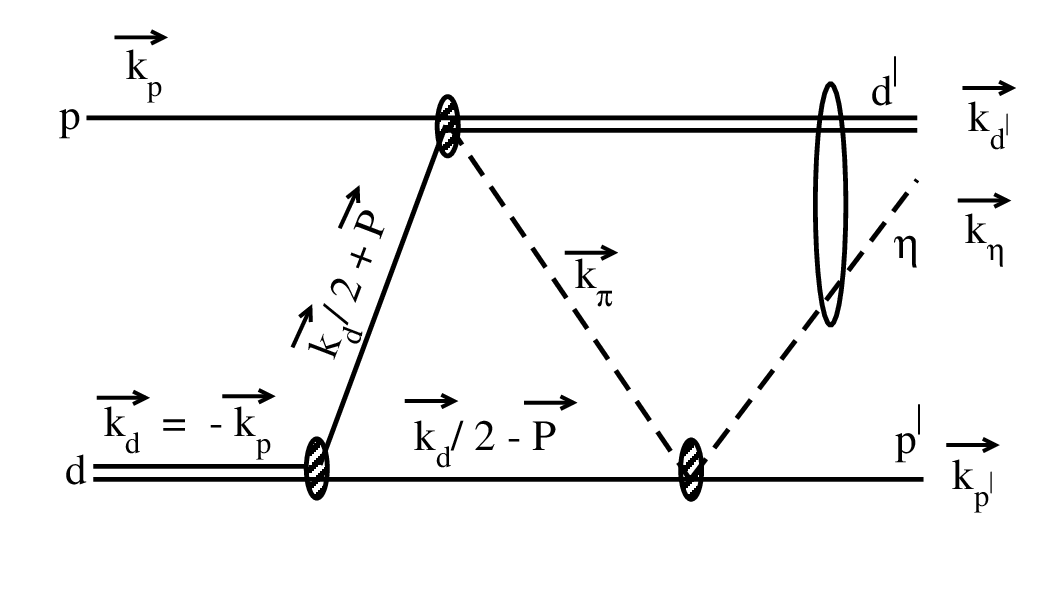}
\caption{\label{fig:eps1}The two step process  production mechanism for the
$p\, d \,\rightarrow \,p \,d \,\eta$ reaction.}
\end{figure}

\subsection{The production mechanism}
For evaluating the $\eta $ production $T$-matrix, 
$\langle\,|\,T_{p d \rightarrow p d \eta}\,|\,\rangle$, we assume a two-step 
mechanism as shown in Fig. 1. In this model, the incident proton produces 
a pion in the first step on interacting with one of the nucleons of the 
target deuteron. In the second step this pion produces an $\eta$ meson on 
interacting with the other nucleon. Both these nucleons are off-shell and 
have a momentum distribution given by the deuteron bound state wave function. 
To write the production matrix, we resort to certain standard approximations
used in literature \cite{kondu3}
(in particular for the triangle diagram appearing in
Fig. 1). The amplitude for the $p N \rightarrow 
\pi d$ process, which in principal is off-shell, is taken at an on-shell 
energy. Considering the high proton beam energy, off-shell effects are not 
expected to be significant. The $\pi \,N\,\rightarrow \,\eta \,N$ process 
is included via an off-shell $T$-matrix.

The production matrix is written as \cite{lage,we3},
\begin{equation}\label{twostep}
\langle \,|\,T_{p\,d\,\rightarrow\,p\,d\,\eta}\,|\,\rangle\,=\,
{3 \over 2}\,i\,\sum_{m's}\,\int 
{d\vec{P}\over (2 \pi)^3}\,\langle\,p\,n\,|\,d\,\rangle \,
\langle\,|\,T_{p\,p\,\rightarrow\,\pi^+\,d} \,|\,\rangle {1 \over 
k_{\pi}^2 \,-\, m_{\pi}^2\,+\,i\epsilon } \langle\,|\,T_{\pi^+\,n\,
\rightarrow\,\eta\,p} \,|\,\rangle
\end{equation}
where, the squared four momentum of the intermediate pion,  
$k^2_{\pi} = E_{\pi}^2 - \vec{k}_{\pi}^2$, with the energy,  
$E_{\pi}$, calculated at zero fermi momentum and 
$\vec{k}_{\pi} = \vec{k}_{\eta} \,+\, 
\vec{k}_{p'} \,-\, \vec{k}_d/2 \,+\,\vec{P}$. 
The summation is over internal spin projections 
and the matrix element $\langle\,p\,n\,|\,d\,\rangle$
represents the deuteron wave function in momentum space, which has been 
written using the Paris parametrization \cite{paris}. The factor $3/2$ 
is a result of summing the diagrams with an intermediate $\pi^0$ and $\pi^+$.

The integral over the pion momentum in above includes the contribution from 
the pole as well as the principal value term. For the pion
propagator itself, as we see, we have taken the plane wave propagator. This
thus excludes any effect in our results due to medium modification of this 
propagator due to other nucleons. This aspect may be worth investigating 
in future.

The $T$-matrix for the intermediate $p\,p\,\rightarrow\,\pi^+\,d$ process has 
been taken from an energy dependent partial wave analysis of the $\pi^+\,d\, 
\rightarrow\,p\,p$ reaction from threshold to $500$ MeV \cite{arndt}. The 
various observables in \cite{arndt} are given in terms of amplitudes which 
are parametrized to fit the existing database. We refer the reader
to \cite{arndt} and the references therein for the relevant expressions
of the helicity and partial wave amplitudes and the notation 
followed by the authors in \cite{arndt}.  

For the $\pi^+\, n\, \rightarrow\, \eta\, p$ sub-process, 
different forms of $T$-matrices are available. We use the $T$-matrix from 
\cite{bhale} which treats the $\pi\,N\,,\,\eta\,N$, and $\pi\,\Delta $ channels 
in a coupled channel formalism. This $T$-matrix consists of 
the meson - N$^*$ vertices and the N$^*$ propagator as given below:
\begin{equation}
T_{\pi^+\,n\,\rightarrow\,\eta\,p}(k^\prime\,,k\,;z)\,=\, 
{g_{N^*}\,\beta^2 \over (k^{\prime\,2}\,+\,\beta^2)}\,\tau_{N^*}(z)\,
{g_{N^*}\,\beta^2 \over (k^2\,+\,\beta^2)}
\end{equation}
with, 
\begin{equation}
\nonumber
\tau_{N^*}(z)\,=\,(\,z\,-\,M_0\,-\,\Sigma_\pi(z)\,-\,\Sigma_\eta(z)\,+\,i\,
\epsilon)^{-1}
\end{equation}
where $\Sigma_\alpha(z)$ ($\alpha\,=\,\pi\,,\,\eta$) are the self energies 
from the $\pi\,N$ and $\eta\,N$ loops. The parameters of this model are, 
$g_{N^*}$ = 0.616, $\beta$ = 2.36 fm$^{-1}$ and $M_0$ = 1608.1 MeV. 
This $T$-matrix reproduces the data on the 
$\pi^+ n \rightarrow \eta p$ reaction very well. 
\begin{figure}
\includegraphics[width =9cm]{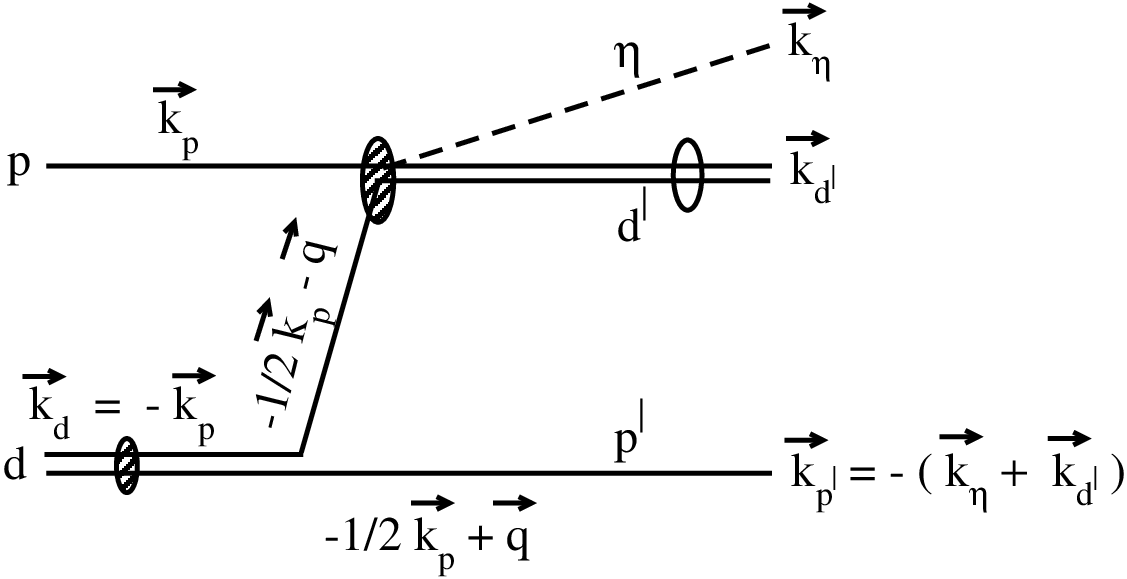}
\caption{\label{fig:eps2}The direct process production mechanism for the
$p\, d \,\rightarrow \,p \,d \,\eta$ reaction.}
\end{figure}

Although the contribution of the direct mechanism (Fig. 2) 
is known to be small (owing to the large momentum transfer
involved in the process) \cite{lage,kond,kama},
for completeness, we calculate its contribution to the total 
cross section. 
The $T$-matrix  for this mechanism can be written as
\begin{eqnarray}\label{rel}
{1 \over 6}\langle 
|T_{p\,d\,\rightarrow\,p\,d\,\eta}|^2\rangle =
{1 \over 4}\langle |T_{p\,n\,\rightarrow\,d\,\eta} (\sqrt{s_{\eta\,d}})|^2
\rangle \times |\,\phi_d(q)|^2,
\end{eqnarray}
where $\phi_d$ represents the deuteron wave function in the initial state.
The spin summed  $\langle |\,T_{p\,n\,\rightarrow\,d\,\eta}\,|^2\rangle$ 
is given in terms of the total cross section for the $p\,n\,\rightarrow\,
d\,\eta$ reaction by
\begin{eqnarray}\label{calen}
\sigma_T(p\,n\,\rightarrow\,d\,\eta)\,=\,
{2\,m_p\,m_n\,m_d \over \pi\,s}\,{|\vec{p_f}| \over |\vec{p_i}|}\,\,
{1 \over 4}\langle |\,
T_{p\,n\,\rightarrow\,d\,\eta}|^2\,\rangle ,
\end{eqnarray}
where $\vec{p_i}$ and  $\vec{p_f}$ are the initial and final momenta c.m. 
system. The momentum transfer $\vec{q}$, as shown in Fig. 2, is defined as,
\begin{equation}
\vec{q}\, = \,{1 \over 2}\, \vec{k_p} \,+ \,\vec{k_{p^\prime}}\, .
\end{equation}
The total cross section, $\sigma_T$, for $p\,n\,\rightarrow\,d\,\eta$ reaction
is taken from the experiments \cite{calen}.\\

\subsection{Final state interaction}
\subsubsection{\bf {$\eta -d$ interaction}}
This is incorporated through a half-off-shell $\eta-d$ $T$-matrix. 
We construct this $T$-matrix using the following two prescriptions:
\begin{enumerate}
\item {\it Factorized form of $T_{\eta d}$}\\
In one ansatz we obtain it by multiplying the on-shell $\eta -d$ 
$T$-matrix by an off-shell extrapolation factor $g( k^\prime\,,\,k )$. 
Requiring that this $T$-matrix goes to its on-shell value in the case 
of on-shell momenta, we write
\begin{equation}
T_{\eta-d}(k\,,\,E(k_0)\,,\,k^\prime)\,=\,g(k\,,\,k_0)\,T_{\eta-d}
(E(k_0))\,g(k^\prime\,,\,k_0),
\end{equation}
with $g(p\,,\,q)\rightarrow 1$ as $p\rightarrow q$. For a half-off-shell 
case, this obviously is the ratio of the half-off-shell to the on-shell 
scattering amplitude.

For the on-shell $\eta -d$ $T$-matrix we use the effective range expansion 
of the scattering amplitude up to the fourth power in momentum, 
 \begin{equation}\label{fsi}
F(k) = \left [{1 \over A}\,+\,{1 \over 2} Rk^2\,+\, Sk^4\,-\,ik \right ]^{-1},
\end{equation}  
where F is related to T by
\begin{equation}
 T_{\eta d}(k,k^\prime)=-\frac{1}{(2\pi)^2\mu _{\eta d}}F_{\eta d}(k,E(k),
k^\prime).
\end{equation}
The effective range expansion parameters (A,R,S) are taken from a recent   
relativistic Faddeev equation (RFE) calculation of  \cite{gar}. 
This calculation uses the relativistic version of the Faddeev equations 
for a three particle $mNN$ system, where m is a meson and it can be an 
$\eta ,\pi $ or a $\sigma$ meson. 
These particles interact pairwise, and these interactions are represented 
with separable potentials. The parameters of the 
$\eta N-\pi N-\sigma N$ potentials are fitted to the $S_{11}$ resonant 
amplitude and the $\pi^-p\rightarrow \eta n$ cross sections. 
The $\eta -d$ effective range parameters obtained from these calculations 
are listed in \cite{gar} for different sets of the meson-nucleon potentials. 
Each of these sets gives a specific value of the $\eta-N$ scattering length, 
which is also listed in \cite{gar}. 

Since the half-off-shell extrapolation factor $g( k^\prime\,,\,k_0)$ 
is not known with any certainty, we choose the following two forms for it.  
\\
(i) Following the method in \cite{wycech} for the 
final state interaction in the $\eta-d$ system, we express the off-shell 
form factor in terms of the deuteron form factor,
\begin{equation}
g(k^\prime\,,\,k_0)= \int d\vec{r}j_0(rk^{\prime}/2)\phi _d^2(r)j_0(rk_0/2)
\end{equation} 
where for the deuteron wave function, $\phi _d(r)$, we take the Paris 
parametrization.\\
(ii) As a second choice, the form factor is taken to be the ratio of the 
off-shell $\eta -d$ $T$-matrix to its on-shell value, where both of them are 
calculated using the three body equations within FRA. The input to 
these calculations is the elementary $\eta-N$ scattering matrix, 
the details of which are given in the next Section.
\item {\it Few body equations within the finite rank approximation} 
\\
The other prescription of $\eta-d$ FSI involves the use of the half-off-shell 
$\eta-d$ $T$-matrix obtained by solving few body equations within the finite 
rank approximation (FRA). For the details of this formalism and the 
expression for the $\eta$-nucleus $T$-matrix, we refer the reader to our 
earlier works \cite{we3}. To mention briefly, the FRA involves restricting 
the spectral decomposition of 
the nuclear Hamiltonian in the intermediate state to the ground state, 
neglecting thereby all excited and break-up channels of the nucleus. 
This is justified in the $\eta-\,^4$He and possibly in the 
$\eta-\,^3$He case, but in $\eta-$deuteron collisions, 
where the break-up energy is just 2.225 MeV, the 
applicability of the FRA may be limited. However, it should be noted that
a comparative study \cite{shev1} of the $\eta -d$ scattering lengths 
calculated using the FRA and the exact Alt-Grassberger-Sandhas 
(AGS) \cite{ags} equations (which include these intermediate excitations) 
shows that they are not very different if the real part of the 
$\eta-N$ scattering length is restricted up to about 0.5 fm.
\end{enumerate}

\subsubsection{\bf {$p-d$ interaction}}
We incorporate the $p-d$ FSI in our calculations 
by multiplying our model $T$-matrix by the inverse Jost function, 
$[J(p)]^{-1}$. We include the FSI in both the 1/2 and 3/2 spin states 
of $p-d$ and restrict it to the s-wave.
Since the $p$ and $d$ are charged we 
also include the Coulomb effects. Following standard 
procedure, we write the Jost function in terms of phase shifts and
use the effective range expansion for the later.

The complete expression for the s-wave inverse Jost function squared  
is written as,  
\begin{equation}\label{pdj}
[J_o\,(k_{pd})]^{-2}\,=\,[J_o\,(k_{pd})]^{-2}_Q\,+\,
[(1\,+\,{|E_B| \over E})J_o\,
(k_{pd})]^{-2}_D .
\end{equation}
Here, to include the effect of the existence of one bound state, namely, 
the spin 1/2 state ($^3He$), the doublet Jost function is multiplied by a 
factor $(1\,+\,{|E_B| \over E})$, where $|E_B|$ is the separation energy 
of $^3He$ into $p-d$. Its value is taken to be 5.48 MeV.
  
The expressions for spin quadruplet 
(Q) and doublet (D) $[J_o\,(k_{pd})]^{-2}$ are given by 
\begin{equation}\label{quad}
[J_o\,(k_{pd})]^{-2}_Q\,=\, {(k_{pd}^2\,+\,\alpha^2)^2\,(b^c_Q)^2 \over 4}\,
\times {1 \over 3\,C_o^2\,k_{pd}^2}\,\left[{2 \over 1\,+\,cot^2\,\delta_Q}\,
\right]
\end{equation}
\begin{equation}\label{dup}
[J_o\,(k_{pd})]^{-2}_D\,=\, {(k_{pd}^2\,+\,\alpha^2)^2\,(b^c_D)^2 \over 4}\,
\times {1 \over 3\,C_o^2\,k_{pd}^2}\,\left[{1 \over 1\,+\,cot^2\,\delta_D}\,
\right]
\end{equation}
where,
\begin{eqnarray}
\alpha&=& \left(\,{1 \over b^c_{\mu}}\,\right)\left[1\,+\,
\left(\,1\,+\,{2\,b^c_{\mu} \over a^c_{\mu}}\,\right)^{1 \over 2}\right]
\end{eqnarray}
and $a^c_{\mu}$ and $b^c_{\mu}$ are defined as
\begin{eqnarray}
{1 \over a^c_{\mu}}&=&{1 \over C_o^2}\left[\,{1 \over a_{\mu}}\,-\,
2\,\gamma\,k_{pd}\,H_{\gamma}\,\right]
\\
b^c_{\mu}&=&{b_{\mu} \over C_o^2}
\end{eqnarray}
where $\mu$ stands for either Q or D.
The factor $C_o^2$ in above has its origin in the Coulomb interaction.
The phase shifts $\delta_{Q\,,\,D}$ are obtained from an effective-range 
expansion \cite{bethe,meyer},
\begin{equation}\label{pde}
C_o^2\,k_{pd}\,cot\,\delta_{\mu}\,=\,-{1 \over a_{\mu}}\,+\,{1 \over 2}
\,b_{\mu}\,k_{pd}^2\,-\,2\,\gamma\,k_{pd}\,H_{\gamma}
\end{equation}
\begin{eqnarray}
\gamma&=&{\alpha\,m_{red} \over \hbar\,k_{pd}}
\\
C_o^2&=&{2\,\pi\,\gamma \over e^{2\,\pi\,\gamma}\,-\,1}
\\
H_{\gamma}&=&\sum_{n\,=\,1}^\infty{\gamma^2 \over n(n^2\,+\,
\gamma^2)}\,-\,ln\,(\gamma)\,-\,0.57722
\end{eqnarray}
Here $m_{red}$ is the reduced mass in the $p-d$ system, $\gamma$ the
Coulomb parameter and $\alpha $ is the usual 
electromagnetic coupling constant. The values of the expansion 
coefficients $a_{\mu}$, $b_{\mu}$ in Eq. (\ref{pde}) are taken as 
$a_Q$ = 11.88 fm, $b_Q$ = 2.63 fm, $a_D$ = 2.73 fm, and $b_D$ = 2.27 fm. 
They have been determined from a fit to the $p-d$ elastic-scattering 
phase shifts in the relative $p-d$ momentum range up to around 200 MeV/c 
\cite{arv}.

The above expression for the Jost function has the required property that 
for large $p$, $J_0(p) \rightarrow 1$.

\section{Results and Discussion}
Before we discuss the results of the present work, in order to highlight 
the FSI effects in the experimental $\eta -d$ invariant mass distribution 
we remove the phase space from the experimental $d\sigma/dM_{\eta d}$ and 
plot in Fig. 3 the $|f|^2$, which is then given by,
\begin{equation}\label{fsq}
|f|^2\,=\,{d\sigma\over d M_{\eta d}}\,\cdot\,{1 \over {\rm phase\,\,space}},
\end{equation}
where,
\begin{equation}\label{phase}
{\rm phase\,\,space}\,=\,{m_p^2\,m_d^2 \over 12\,(2\,\pi)^5\,s\,
|\vec{k_p}|}\,\,\int d\Omega_{p^\prime}\,\,|\vec{k_{p^\prime}}|\,\,
|\vec{k}_{\eta\,d}|\,\,d\Omega_{\eta\,d}
\end{equation}
as a function of the excess energy, $Q_{\eta d}\,=\,M_{\eta d}\,-\,m_{\eta}
\,-\,m_d$, where $M_{\eta d}$ is the invariant mass of the $\eta-d$ system. 
In this figure we also show the plane wave result (i.e. $T_{p\,d\,\rightarrow
\,p\,d\,\eta}$ does not include any FSI). The cross section, 
$d\sigma / d M_{\eta\,d}$ in Eq. (\ref{fsq}), is evaluated for each 
$M_{\eta d}$ by performing an integral over the $p-d$ centre of mass 
momenta, $k_{pd}$. The range of the allowed values of $k_{pd}$ at 
each $M_{\eta d}$ is shown by the hashed region. One clearly sees a 
large enhancement in the experimental $|f|^2$ near small values of 
$Q_{\eta\,d}$, which, most likely is due to the $\eta-d$ FSI. 
We also observe a rise at large 
values of $Q_{\eta d}$. Examining the range of $p-d$ relative momenta which 
contribute to $|f|^2$ at each $Q_{\eta d}$, one can see that this rise occurs 
at small values of $k_{pd}$, indicating thereby the possibility of a large 
effect of $p-d$ FSI in this region.  

\begin{figure}
\includegraphics[width = 9cm]{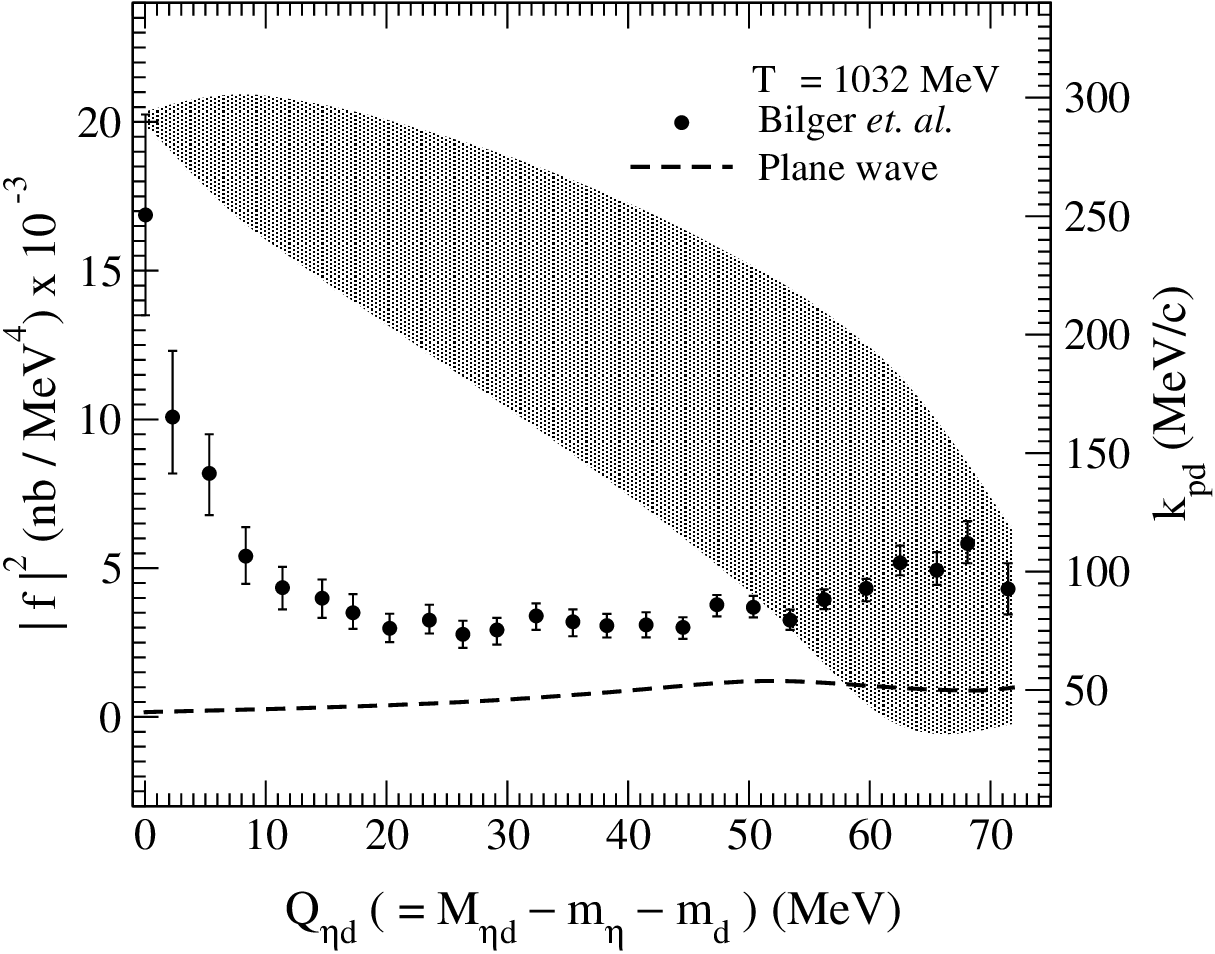}
\caption{\label{fig:eps3}The ratio of experimental differential 
cross sections \cite{bilger} to the phase space (Eq. (\ref{phase})) 
as a function of the excess energy, 
$Q_{\eta d}$, along with range of $p-d$ relative momenta, $k_{pd}$ 
(hashed region), contributing to $|f|^2$ at each $Q_{\eta d}$.}
\end{figure}  

In Fig. 4, we show two sets of the calculated $|f|^2$ along with the 
experimental results for a beam energy of 1032 MeV. These results include 
only $\eta -d$ FSI. We limit the range of $Q_{\eta d}$ up to about 10 MeV,  
where, this effect is large. In Fig. 4(a) we show results for the factorized 
prescription with the off-shell factor generated from the deuteron form 
factor and the on-shell part arising from the relativistic Faddeev 
equation (RFE) calculation of \cite{gar}. The results are shown for  
three different sets of interaction parameters in the RFE. 
Since these sets give uniquely different values of the $\eta-N$ scattering 
lengths $a_{\eta N}$, we identify them by their corresponding  
$a_{\eta N}$ values. For the results presented here, these values are 
0.42 + i0.34 fm, 0.75 + i0.27 fm and 1.07 + i0.26 fm. We see that 
our results reproduce the enhancement seen in the experimental $|f|^2$ 
at small values of $Q_{\eta d}$. The absolute magnitude depends upon the 
choice of the RFE parameters. It increases with $a_{\eta N}$, which 
designate these parameter sets. The set corresponding to 
$a_{\eta N} = 1.07 + i0.26$ fm, gives results closest to the experimental 
values.  
\begin{figure}
\includegraphics[width =8cm]{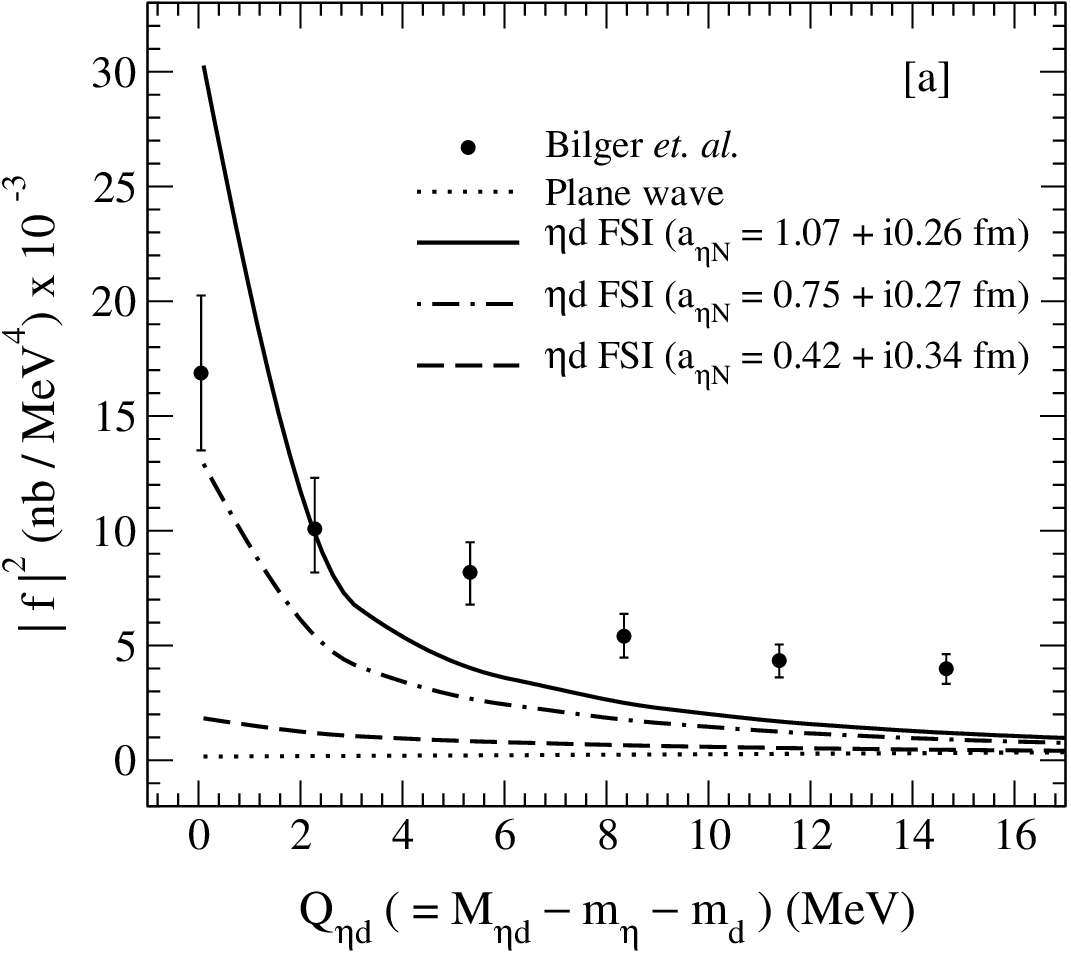}
\includegraphics[width =8cm]{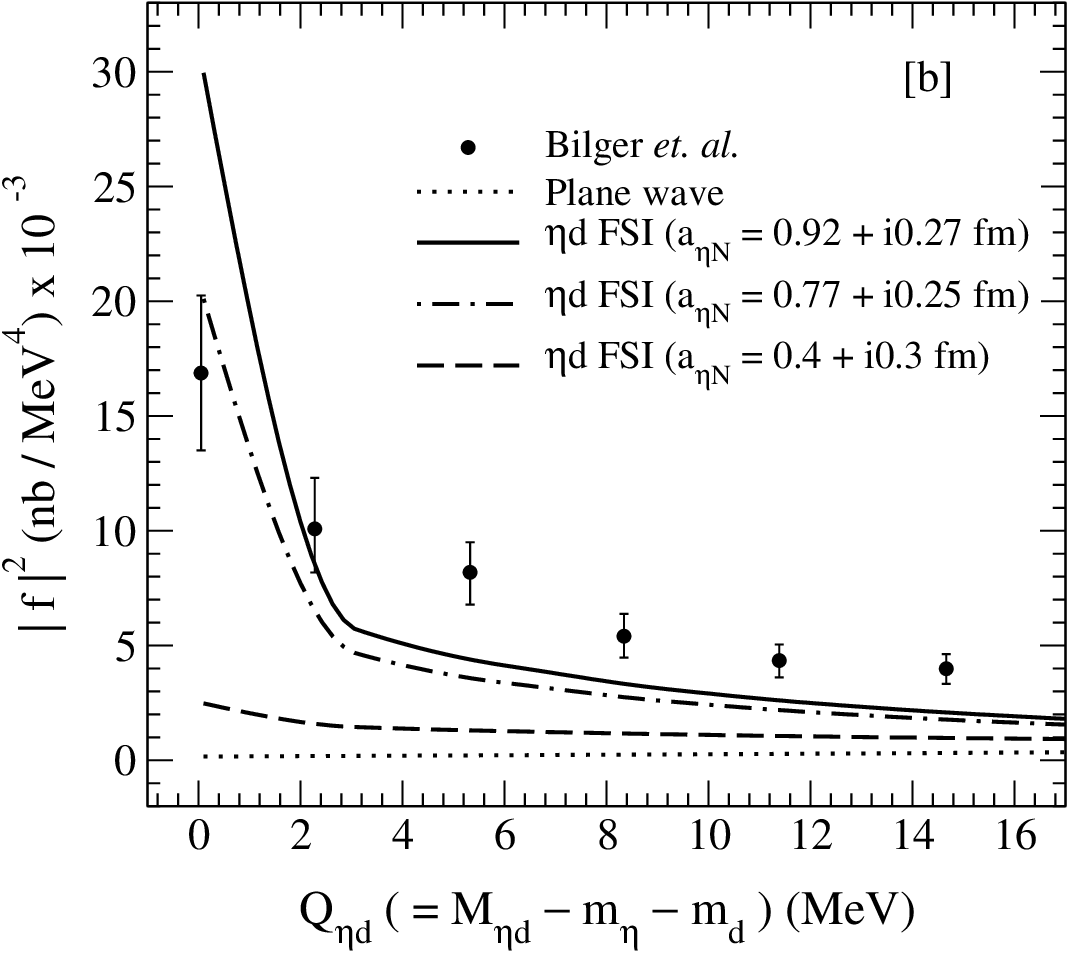}
\caption{The calculated $|f|^2$ along with the experimental 
results for a beam energy of 1032 MeV. (a) The results correspond 
to the factorized form of $T_{\eta d}$ with the off-shell factor generated 
from the deuteron form factor. (b) The results correspond to 
$T_{\eta d}$ obtained from few body equations within the FRA. 
The data is the same as in Fig. 3.}
\end{figure}

In Fig. 4(b) we show $|f|^2$ calculated using few body equations within 
the FRA, for $\eta-d$ FSI. These results are shown for 
three different inputs of the $\eta-N$ $T$-matrix taken from \cite{green}. 
The choice of these $T$-matrices is such that their scattering 
length values are close to those used in Fig. 4(a). Though this model has 
the limitation of retaining the intermediate 
nucleus in its ground state in the $\eta-$ nucleus elastic scattering, 
the off-shell re-scattering effects have been properly included. If we 
compare Fig. 4(a) and 4(b), the two sets of results are similar. 
\begin{figure}
\includegraphics[width =8cm]{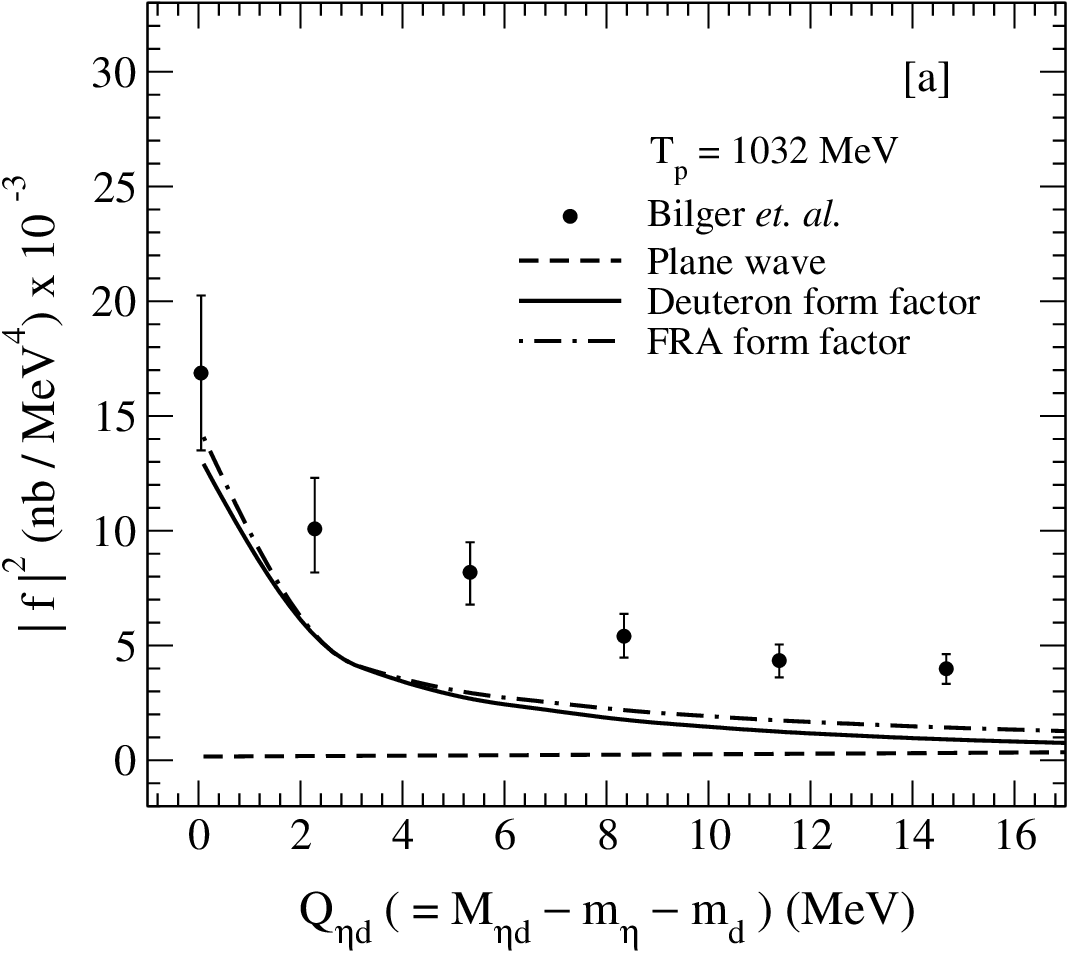}
\includegraphics[width =8.07cm]{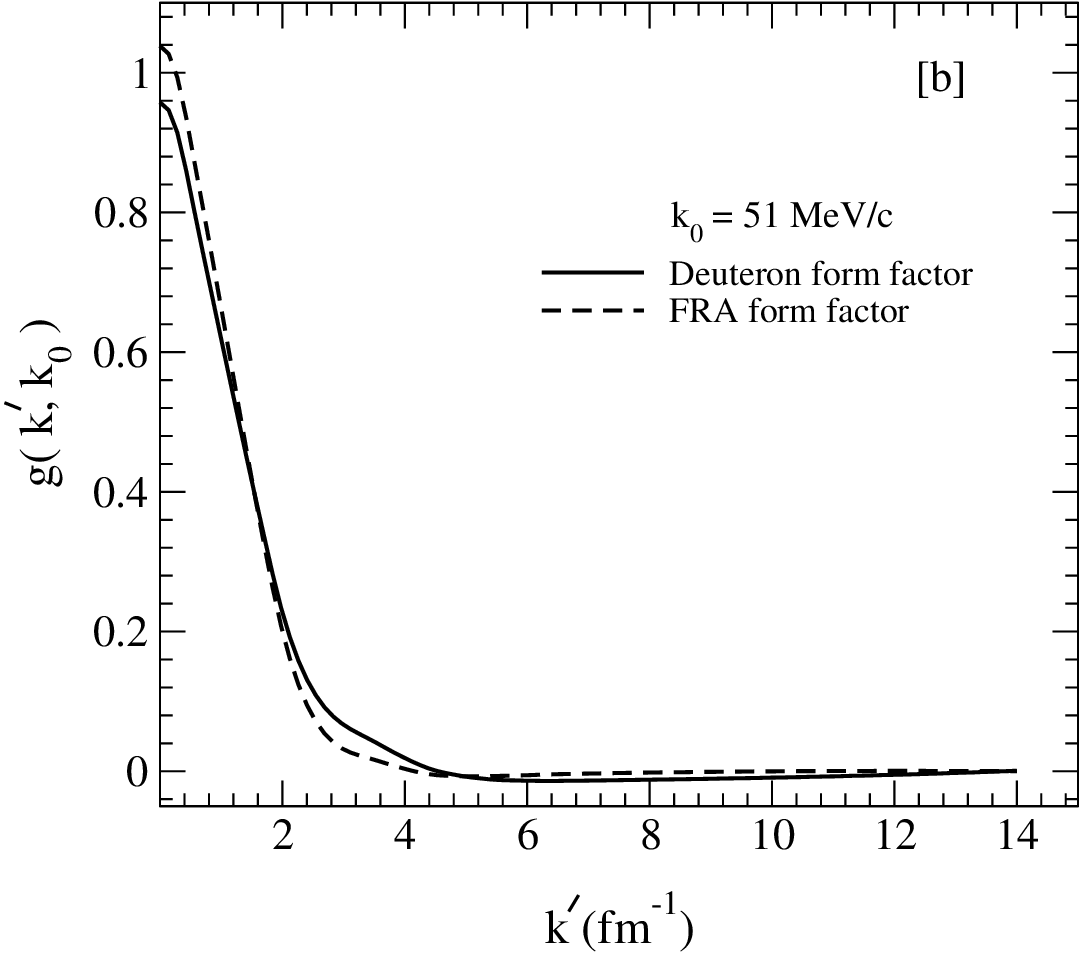}
\caption{Comparison of the two form factors for $a_{\eta\,N}$
 = 0.75 + i0.27 fm. (a) Effect of using two different off-shell extrapolation 
factors for $\eta-d$ FSI on $|f|^2$. (b) Two form factors as function of 
off-shell momentum ($k^{\prime}$).}
\end{figure}

In order to check the sensitivity of the results to the off-shell form factor
used in the factorized $\eta-d$ $T$-matrix, 
in Fig. 5(a), we show the $|f|^2$ calculated using two different off-shell 
form factors. The on-shell $T_{\eta d}$ is obtained from RFE and the 
off-shell part is 
either treated with a deuteron form factor (solid line) or a few body FRA 
form factor (dash dotted line) as explained in section II B. 
The elementary $\eta-N$ $T$-matrix parameters required for the calculation 
of the FRA form factor are taken from the 
parametrization of Green and Wycech \cite{green}. Even though the results 
(as shown in Fig. 4(a)) corresponding to the a$_{\eta\,N}$ = 1.07 + i0.26 fm
seem to be the closest to the data, to compare the effect of using different 
off-shell form factor, we choose the results corresponding to
a$_{\eta\,N}$ = 0.75 + i0.27 fm. We make this choice such that we can compare
the two calculations for the inputs corresponding to a similar $\eta-N$ 
scattering length. It should be expected then, that the off-shell form factors 
obtained from two different methods should not differ much. 
This is seen explicitly in Fig. 5(b) where the two form factors are shown 
as a function of off-shell momentum ($k^\prime$) for an on-shell value, $k_0$ 
near the low energy peak in the 
$\eta-d$ invariant mass distribution (to be discussed in Fig. 9 later). 

Next, we include in our calculations the effect of the $p-d$ FSI. This is 
done by multiplying the $p d \rightarrow p d \eta$ squared $T$-matrix  
(Eq. (\ref{tfull})) used above by the inverse Jost function squared in 
Eq. (\ref{quad}) and Eq. (\ref{dup}), and integrating it over the allowed 
range (as shown in Fig. 3) of $p-d$ momenta, $k_{pd}$ for each $Q_{\eta\,d}$. 
We show these results in Fig. 6 for the RFE (with deuteron form factor) 
model of $\eta-d$ FSI, for the parameter set corresponding to  
a$_{\eta\,N}$ = 1.07 + i0.26 fm. We find, that the $p-d$ FSI affects the 
results in the whole region of $Q_{\eta\,d}$, while the effect of $\eta-d$
FSI is confined to small value of $Q_{\eta d}$. The large effect of $p-d$ 
FSI in the region of small $Q_{\eta d}$, however, may not be taken with 
confidence as the value of $k_{pd}$ in this region is large (as shown in
Fig. 3), where the s-wave effective range expansion (Eq. (\ref{pde})) for 
the calculation of Jost function might not be sufficient. In any case, 
it appears that the effects of both the $\eta-d$ and the $p-d$ FSI on the 
$\eta-d$ invariant mass distribution are significant. If we disregard the 
calculated $p-d$ effect for small $Q_{\eta d}$, the $\eta-d$ and $p-d$ FSI
dominate in regions well separated from each other.
\begin{figure}
\includegraphics[width =8cm]{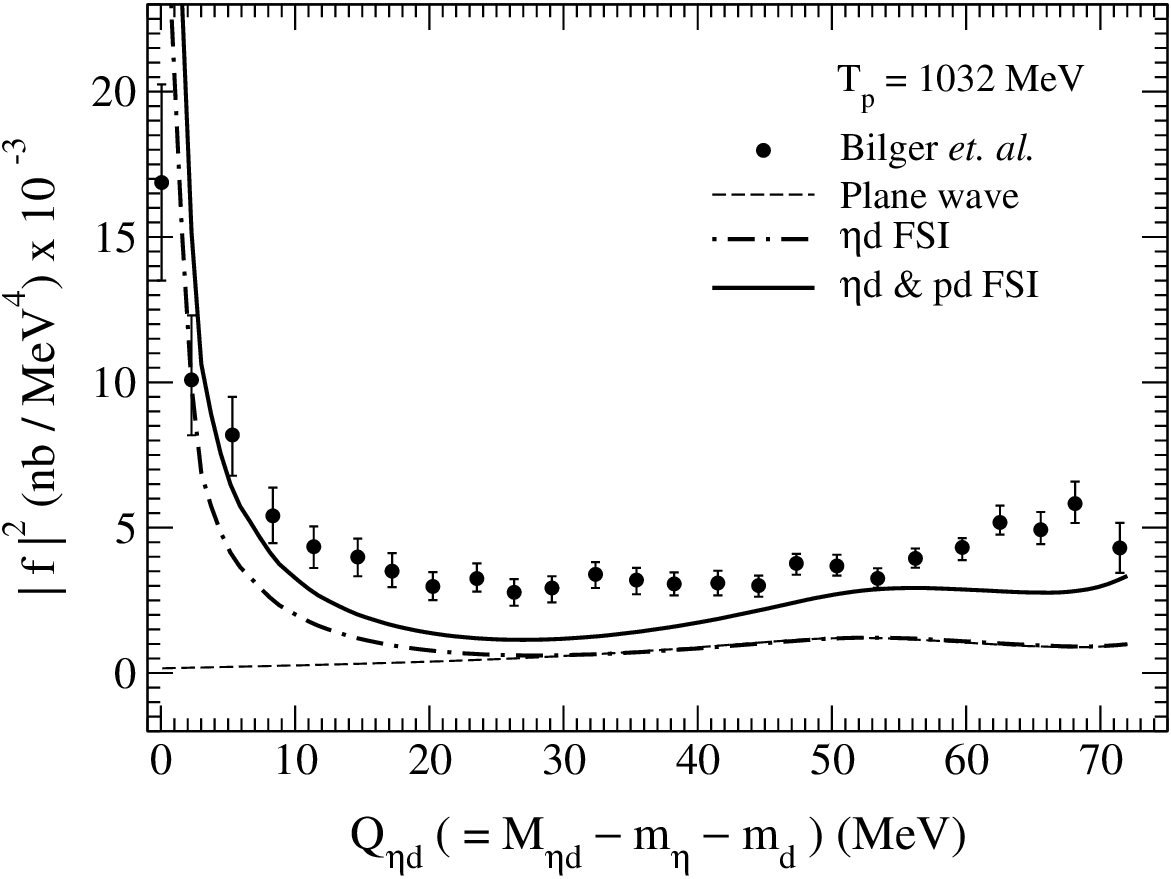}
\caption{\label{fig:eps6}The proton-deuteron final state interaction
effects on the $p\,d\,\rightarrow\,p\,d\,\eta$ reaction at the beam 
energy of 1032 MeV. The dashed line shows the plane wave results and 
the dashed dot (solid) line shows the effect of $\eta-d$ ($\eta-d$ \& 
$p-d$) FSI for $a_{\eta\,N}$ = 1.07 + i0.26 fm.}
\end{figure}

\begin{figure}
\includegraphics[width =9cm]{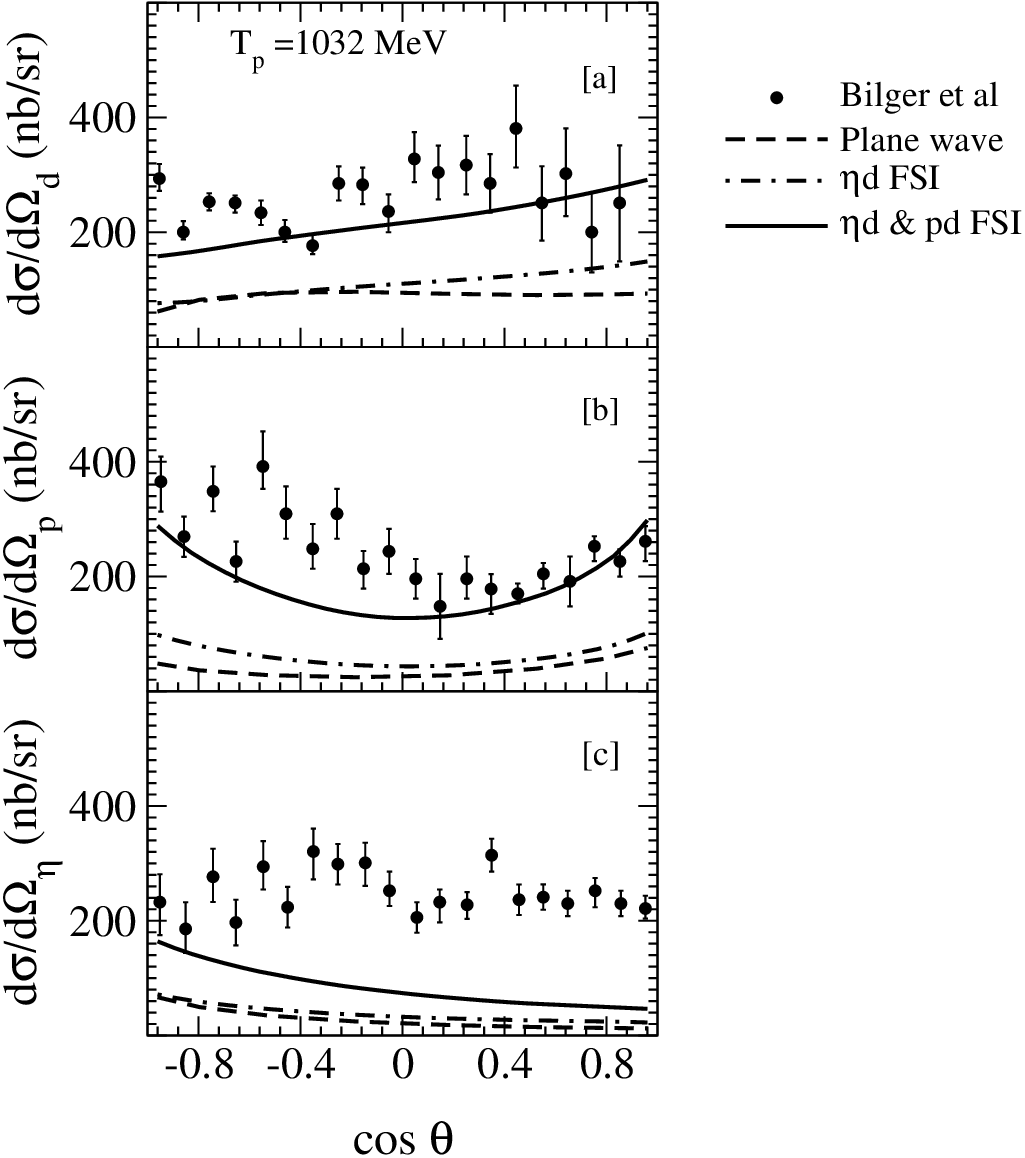}
\caption{\label{fig:eps7}The calculated angular distributions of (a) the 
deuteron, (b) the proton and (c) the $\eta$, along with the measured cross 
sections for $a_{\eta\,N}$ = 1.07 + i0.26 fm \cite{bilger}.}
\end{figure} 

Apart from the FSI, another important ingredient of our calculations is 
the two-step 
description of the production vertex. Because of the large momentum 
transfer, we believe, as has also been stressed in Ref. \cite{wilkin}, that 
the angular distribution of the outgoing particles is 
probably more sensitive to the description of the production vertex.
Inclusive angular distributions have been 
measured for all the three outgoing particles in the 
$p\,d\,\rightarrow\,p\,d\,\eta $ reaction. In Fig. 7, we show the 
calculated angular distributions for all the 
three outgoing particles along with the measured distributions.
We show results without any FSI, with $\eta-d$ FSI and with $\eta-d$ and 
$p-d$ FSI both included.
As each angle has contribution from a range of $Q_{\eta\,d}$ as well as 
$k_{pd}$, the calculated results include integration of the cross section 
over these variables. 
We find that the observed nearly isotropic nature of the experimental 
angular distributions for the proton, deuteron and eta already gets 
reproduced by the plane wave calculations. The effect of both 
$\eta-d$ and $p-d$ FSI is large and persists over all the angles. Their 
inclusion brings the magnitudes of the proton and deuteron angular 
distributions near to experiments. The magnitude of the eta distribution, 
however, does not seem to be affected much with the FSI. 
\begin{figure}
\includegraphics[width =8cm]{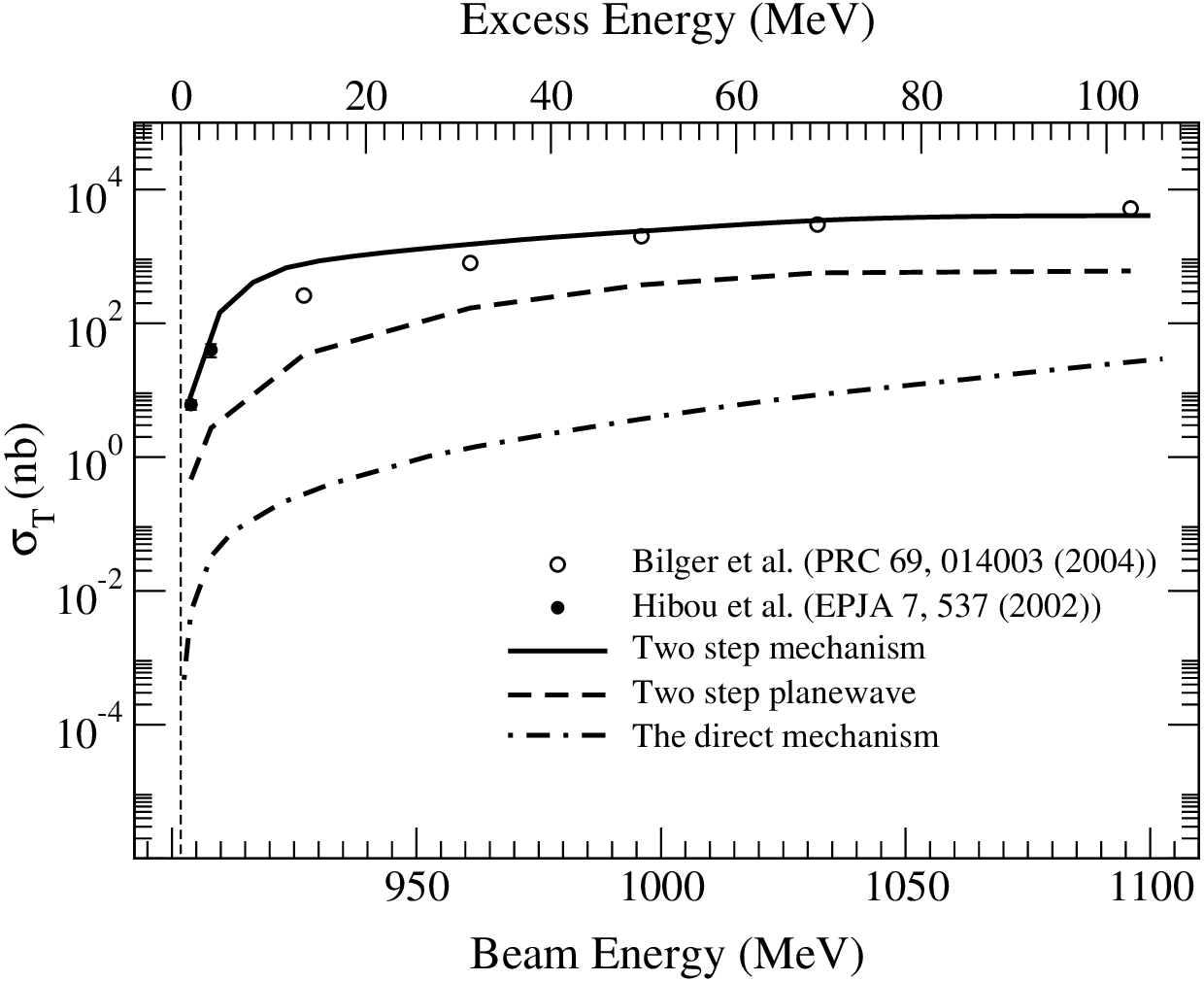}
\caption{\label{fig:eps8}A comparison of the total cross section for the 
$p\,d\,\rightarrow\,p\,d\,\eta$ reaction calculated with the description
of the production vertex as a two step mechanism and direct mechanism, along
with the measured cross sections for $a_{\eta\,N}$ = 1.07 + i0.26 fm 
\cite{bilger,hibou}.}
\end{figure}

Experimental data also exist on the total cross section. In Fig. 8, 
we compare the total cross sections calculated including both 
the $\eta-d$ and $p-d$ FSI with the measured 
cross sections. The results are shown with the factorized form of $\eta-d$ FSI 
with deuteron form factor for the set corresponding to $\eta-N$ scattering
length equal to 
$1.07 + i0.26$ fm. As we see, the calculated cross sections are  
in good agreement with the experimental data. 

In Fig. 8 we also give the cross sections calculated for 
the one-step direct mechanism (Fig. 2) mentioned in the previous
section. Near threshold, these cross sections are about four
orders of magnitude below those obtained from the 
two-step model and two orders of magnitude smaller in the high energy range. 
As mentioned in the Introduction, 
this observation is similar to that in other works involving 
large momentum transfer reactions \cite{lage,kond,kama}, and is 
understandable because the momentum 
transfer continues to be large ($\sim 600$ MeV/c) in the $p\,d\,\rightarrow\,
p\,d\,\eta$ reaction even at an excess energy as large as 100 MeV. 
\begin{figure}
\includegraphics[width =8cm]{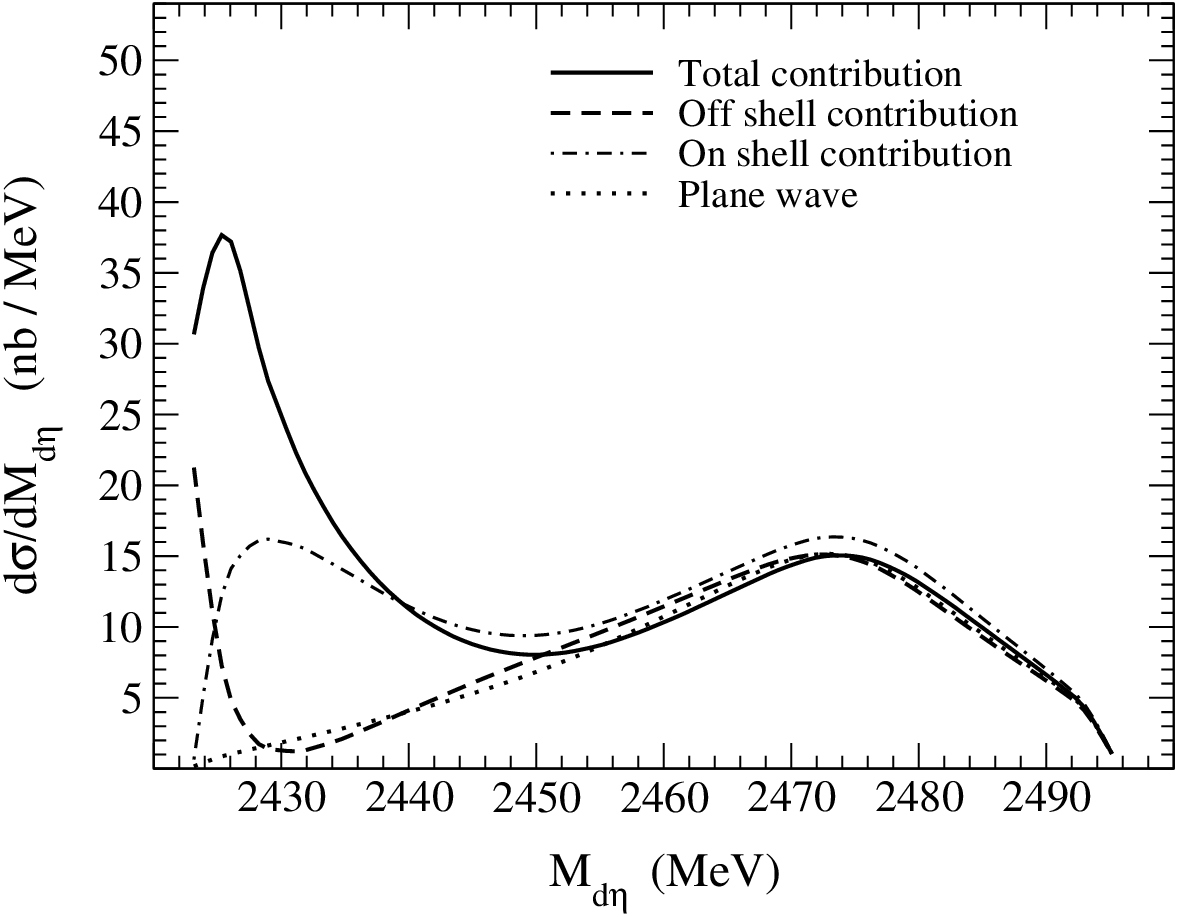}
\caption{\label{fig:eps11}Contributions from the off-shell and the on-shell
$\eta-d$ scattering in the final state. The results are for 
$a_{\eta\,N}$ = 1.07 + i0.26 fm with the inclusion of only the $\eta-d$ FSI.}
\end{figure}

Now we make an observation about the importance of off-shell scattering
in treating $\eta-d$ FSI near threshold. The scattering part of the $\eta-d$ 
wave function (Eq. (\ref{psi})), gets contributions from the off-shell as 
well as the on-shell scattering in the nucleus. To see quantitatively 
the relative importance of these two contributions to the cross section for 
the $p\,d\,\rightarrow\,p\,d\,\eta $ reaction, in Fig. 9 we show their 
contributions separately in the $\eta-d$ invariant mass distribution.
These results include only the $\eta-d$ FSI generated from the factorized
prescription using RFE and the deuteron form factor for the $\eta-d$ 
$T$-matrix. We find that near threshold the off-shell scattering completely 
dominates the threshold enhancement. At higher excess energy, as expected, 
the on-shell contribution takes over. 
\begin{figure}
\includegraphics[width =8cm]{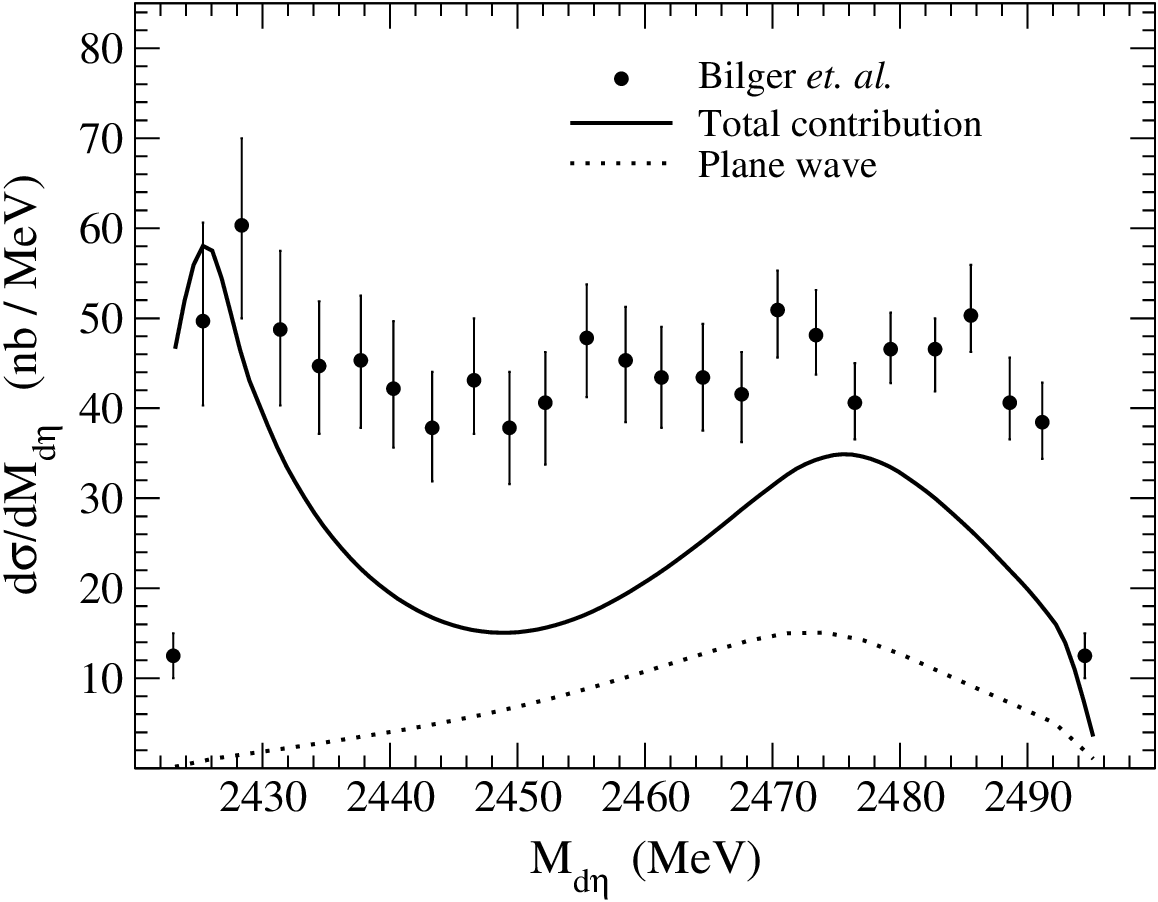}
\caption{\label{fig:eps11}A comparison of the calculated results including
both the $\eta-d$ and $p-d$ FSI with the experimental results. The results 
are for $a_{\eta\,N}$ = 1.07 + i0.26 fm.}
\end{figure}

Finally we show the nature of agreement of our calculated results with the 
invariant $\eta-d$ mass distribution. In Fig. 10, we compare the calculated
results including both the $\eta-d$ and $p-d$ FSI with the experimental
results. The results are for a$_{\eta\,N}$ = 1.07 + i0.26 fm calculated with 
the factorized prescription using RFE and the off-shell factor generated from
the deuteron form factor. As we see the overall agreement is reasonably good.

\section{Summary}
The invariant $\eta-d$ mass distribution in the $p\,d\,\rightarrow 
\,p\,d\,\eta$ reaction has been studied
by describing the production mechanism in terms of 
a two step model with a pion
being produced in the intermediate state. 
The $\eta-d$ final state interaction (FSI) has been included in (a) a 
factorized form involving an on-shell $T_{\eta d}$ and two types of
off-shell form factors and (b) by solving few body equations within the FRA.
The $p-d$ FSI is included through a Jost function. 
The conclusions of this investigation can be summarized as:
\begin{enumerate}
\item Experimentally observed large enhancement in the cross section 
near small $\eta-d$ excess energy, $Q_{\eta d}$ is reproduced by the 
$\eta-d$ FSI. The rise in the cross section at large $Q_{\eta d}$ 
(which corresponds to a range of small momenta, $k_{pd}$) can be 
accounted for by the $p-d$ FSI. 
\item Quantitative reproduction of the large enhancement requires $\eta-d$ 
FSI corresponding to large values of $a_{\eta N}$. In the present 
calculation it is around 1.07 + i0.26 fm.   
\item The calculations successfully reproduce the observed 
isotropic angular distribution of the proton and the 
deuteron in the final state. The total cross sections for the 
$p d \rightarrow p d \eta$ reaction are also well reproduced. 
\item 
The off-shell part of the $\eta-d$ scattering dominates near threshold.
\item The results for two different choices of the off-shell extrapolation 
factor in the factorized form of the $\eta-d$ FSI are similar. 
\end{enumerate}

\section{Acknowledgments}
The authors wish to thank R. A. Arndt for providing the computer codes
for evaluating the $p\,p\,\rightarrow\,\pi^+\,d$ $T$-matrix. 
This work is done under a research grant by the
Department of Science and Technology, Government of India. The 
authors (NJU, KPK and BKJ) gratefully acknowledge the same.


\begin{thebibliography}{99}
\bibitem{bhale}
R. S. Bhalerao and L. C. Liu, Phys. Rev. Lett. {\bf 54} (1985) 865.
\bibitem{hailiu} 
Q. Haider and L. C. Liu, Phys. Lett. B {\bf 172}, 257 (1986); 
{\it ibid} {\bf 174}, 465(E) (1986); {\it ibid}, 
Phys. Rev. C {\bf 66}, 045208 (2002);  
N. G. Kelkar, K. P. Khemchandani and B. K. Jain,
J. Phys. G:Nucl. Part. Phys. {\bf 32}, L19 (2006); arXiv:nucl-th/0601080;
for experimental claims, see  
M. Pfeiffer {\it et. al}., Phys. Rev. Lett. {\bf 92}, 252001 (2004) and 
references therein.
\bibitem{bilger}
R. Bilger {\it et. al.}, Phys. Rev. {\bf C69} (2004) 014003.
\bibitem{we3}
K. P. Khemchandani, N. G. Kelkar and B. K. Jain, Nucl. Phys. {\bf A708}
(2002) 312; {\it ibid}, Phys. Rev. {\bf C 68} (2003) 064610. 
\bibitem{lage}
J. M. Laget and J. F. LeColley, Phys. Rev. Lett. {\bf 61} (1988) 2069.
\bibitem{arndt}
R. A. Arndt, I. Strakovsky, R. L. Workman and D. A. Bugg, Phys. Rev.
{\bf C48} (1993) 1926. The amplitudes can be obtained from the SAID
program available at (http://gwdac.phys.gwu.edu).
\bibitem{gar}
H. Garcilazo, Phys. Rev. {\bf C71} (2005) 048201.
\bibitem{rakit}
S.A. Sofianos and S.A. Rakityansky, nucl-th/9707044,
S. A. Rakityansky {\it et. al.}, Phys. Rev. {\bf C53 } (1996) 2043.
\bibitem{watmig}
K. M. Watson, Phys. Rev. {\bf 88}, 1163 (1952); 
A. B. Migdal, Sov. Phys.-JETP {\bf 1}, 2 (1955).
\bibitem{jost}
J. Gillespie, Final State Interactions, Holden-Day, Inc.,
San Francisco (1964).
\bibitem{gold}
Goldberger and Watson, Collision Theory, John Wiley $\&$ Sons, Inc., New York,
(1964).
\bibitem{shyam}
R. Shyam, Phys. Rev. C {\bf 60}, 055213 (1999).
\bibitem{dks}
J. Dubach, W. M. Kloet and R. R. Silbar, Phys. Rev. C {\bf 33}, 373 (1986).
\bibitem{penya}
H. Garcilazo and M. T. Pe\~na, Phys. Rev. {\bf C72} (2005) 014003 ; 
{\it ibid}, {\bf C66} (2002) 034606.
\bibitem{wycech}
S. Wycech and A. M. Green, Phys. Rev.  {\bf C64} (2001) 045206.
\bibitem{green}
A. M. Green and S. Wycech, Phys. Rev. {\bf C71 } (2005) 014001; see 
(also the erratum in, Phys. Rev. {\bf C72} (2005) 029902(E).
\bibitem{wilkin}
U. Tengblad, G. F$\ddot{a}$ldt and C. Wilkin, Eur. Phys. J. {\bf A25} 
(2005) 267; arXiv:nucl-th/0506024.
\bibitem{kond}L. A. Kondratyuk and Yu. N. Uzikov, 
Phys. Atom. Nucl. {\bf 60} 468 (1997); arXiv:nucl-th/9510010.
\bibitem{kama}
V. I. Komarov, A. V. Lado and Yu. N. Uzikov, J. Phys. G:Nucl. Part. Phys. 
{\bf 21}, L69 (1995); Phys. Atom. Nucl. {\bf 59}, (1996) 804; 
arXiv:nucl-th/9804050.
\bibitem{kondu3}
L.A. Kondratyuk and M.G. Sapozhnikov, Phys. Lett. {\bf B220}, (1989) 333; 
L.A. Kondratyuk, A.V. Lado and Yu.N. Uzikov, 
Phys. Atom. Nucl. {\bf 58}, (1995) 473, Yad. Fiz. {\bf 58}, 1995 (524); 
A. Nakamura and L. Satta, Nucl. Phys. {\bf A445}, (1985) 706; 
V.M. Kolybasov and N.Ya. Smorodinskaya, Phys. Lett. {\bf B37}, (1971) 272.
\bibitem{paris}
M. Lacombe {\it et. al.}, Phys. Lett. {\bf B101} (1981) 139.
\bibitem{calen}
H. Cal$\grave{e}$n Phys. Rev. Lett. {\bf 80} (1998) 2069.
\bibitem{shev1}
N. V. Shevchenko {\it et. al.}, Phys. Rev. {\bf C58 } (1998) 3055(R).
\bibitem{ags}
E. O. Alt, P. Grassberger, W. Sandhas, Nucl. Phys. {\bf B2}
(1967) 167.
\bibitem{bethe}
H. A. Bethe, Phys. Rev. {\bf 76}, 38 (1949).
\bibitem{meyer}
H. O. Meyer and J.A. Niskanen, Phys. Rev. C {\bf 47}, 2474 (1993).
\bibitem{arv}
J. Arvieux, Nucl. Phys. {\bf A221}, 253 (1974).
\bibitem{hibou}
F. Hibou  {\it et. al.}, Eur. Phys. J. {\bf A7 } (2000) 537.
\end{thebibliography}
\end{document}